# Drill-bit-inspired dynamic focal fields for augmented laser materials processing

Evangelos Skoulas[1*], Pinku Yadav[1], Justin Hidjam[1], Aurelien Woher[1], Rainer Kling[2], Beat Neuenschwander[2], Alexander Rack[3], Xavier Maeder[4], Patrik Hoffmann[1], Elia Iseli[1], Sergey Shevchik[1]

1) *EMPA, Laboratory for Advanced Materials Processing, Feuerwerkerstrasse 39, 3602 Thun, Switzerland*
2) *Institute for Applied Laser, Photonics and Surface technologies ALPS, Bern University of Applied Sciences, Pestalozzistrasse 20, 3400 Burgdorf, Switzerland*
3) *ESRF – The European Synchrotron, CS 40220, Grenoble, F-38043, France*
4) *EMPA, Laboratory for Mechanics of Materials & Nanostructures, Feuerwerkerstrasse 39, 3602 Thun, Switzerland*

*Corresponding author: evangelos.skoulas@empa.ch*

## Abstract

Laser manufacturing has advanced through increasingly precise control of power, pulse duration, repetition rate and scan trajectory, yet the spatial intensity profile of the beam is still usually fixed during light–matter interaction. This constraint limits how energy can be delivered to matter, particularly in processes where melt flow, material removal and surface morphology evolve on comparable time and length scales. Here we introduce drill-bit-inspired laser beams that convert the focal intensity distribution from a passive, static spot into an active, programmable processing tool. By combining cylindrical vector beams with rotational vectorial polarization filtering, we create a near diffraction limited two lobe Hermite–Gaussian focus that continuously spins about the propagation axis and can be reconfigured on demand. We establish two operating regimes, dynamic beam spinning and instantaneous beam-profile shifting, and derive closed-form descriptions of the accumulated fluence and effective pulse number governed by the along-scan pitch $\Lambda = u/f$, where $u$ is the scan speed and $f$ is the spin frequency. Across continuous-wave and ultrashort-pulse regimes, this dynamic energy deposition enables low-power metal machining with drilling efficiencies about four times higher than static Gaussian, enhances convective melt flow, promotes pore resorption and reduces retained porosity in keyhole welding, as visualized by in situ X-ray imaging, and turns simple linear scans into programmable surface textures. These results show that dynamic focal-profile control can extend laser processing beyond static beam shaping, opening a broadly applicable route to programmable energy deposition in manufacturing.

# 1. Introduction

From the discovery of the lasing effect to the evolution of modern systems, lasers have emerged as indispensable tools for material processing in many scientific and technological fields. Surface modification[1,2] and coloration[3], subtractive processes such as dicing[4], drilling[5], and additive manufacturing[6], as well as welding[7] and nano-structuring[8], are some of the areas where lasers are extensively used for shaping, reforming, and fabricating materials. To ensure versatility and applicability across different fields and material types, a set of parameters must be carefully adjusted to achieve optimal efficiency and compatibility. The most fundamental parameters include wavelength (λ), power or energy dose (P, E), and exposure time, whether the laser operates in pulsed or continuous wave (CW) mode and scan path. In addition, beam characteristics related to spatial distribution and polarization state, play a critical role[8,9]Until recently, laser materials processing relied on this limited parameters related to energy/power dose values and wavelength.

In recent years, spatial beam shaping has been shown to directly influence the quality of processes such as nanostructure formation[10,11] and additive manufacturing[6,12]. Resent studies use shaped laser beams for steering solidification[13,14], an effect later confirmed in full three-dimensional builds[15] and rationalized by coupled ray-tracing hydrodynamic simulations[16]. In powder bed fusion, Bessel beams suppress keyholing and stabilize the melt pool across a broad parameter space[17], while ring-shaped and top-hat profiles reduce spatter, balling and permit substantially higher build rates[18]. Furthermore high-speed X-ray imaging directly linking the ring profile's manipulation of incident and reflected rays to the elimination of the unstable keyhole tip responsible for porosity[19]. In ultrafast surface structuring, spatially tailored fields vortex and vector beams with engineered polarization distributions, or lossless top-hat lines generated by multi-plane light conversion provide control over the orientation, regularity, and areal throughput of laser-induced periodic surface structures[20,21]. All these examples highlight that the laser beam itself, acting as the tip of the carving drill, plays a decisive role in determining the outcome of laser material processing across diverse applications.

In this work, we present a drill-bit inspired light shaping approach based on rotational vectorial polarization filtering of cylindrical vector beams. Consequently, the laser beam is now not a stable focused energy point but a dynamic steering bit at the focusing diffraction limit. We introduce two operating modes: a pure spinning mode (SM) and a rapid beam profile shifting mode (PSM) that preserves the beam size compared to the standard Gaussian ($\mathrm{LG}_0^0$) output of the source, operating close to the diffraction limit. We demonstrate its performance in three use cases, showing that it enables consistent machining with CW sources and enhances drilling efficiency (drilled area over same power) under identical power and operating conditions, even with variable focusing in thin and bulk metallic substrates. High speed X ray imaging investigation showed that the use of spinning drill bit beams in keyhole mode can create more pronounced convective flow, promotes pore resorption inside the melt pool of highly pore-prone metal targets. Furthermore, spinning beams enable the fabrication of complex surface textures using linear scanning strategies that are fully programmable and can significantly reduce processing time. We believe that this work provides an additional degree of freedom as a crucial parameter for advanced laser material processing. In short, spinning beams combine the advantages of dynamic energy delivery, improved melt physics, and enhanced controllability, making them highly powerful for next-generation laser processing.

# 2. Material and methods

## 2.1 Laser sources

Three laser sources were employed in this study. Two were continuous-wave (CW) ytterbium fiber lasers (IPG Photonics, models YLR-1000 and YLR-50; maximum output powers 1 kW and 50 W, respectively; emission wavelength $\lambda$ = 1070 nm). YLR-1000 served as the processing beam for keyhole-mode melting and the YLR-50 for low-power machining and drilling. For these experiments, beam delivery used a plano-convex lens (f = 150 mm) and sample translation used a 3-axis Zaber motorized stage. The third source was a femtosecond fiber laser (aeroPULSE, NKT Photonics) delivering pulses of 450 fs duration at a central wavelength of 1030 nm and a repetition rate of 103 kHz, used for the surface texturing experiments. Beam delivery and scanning for the femtosecond experiments were performed with a galvanometric scan head (SCANLAB GmbH, hurrySCAN) equipped with an f-theta objective (focal length 150 mm), yielding a focused spot diameter of 68 µm at the $1/e^2$ at the focal plane.

The rotating (“spinning drill-bit”) beam configuration was realized using a free-space optical arrangement, producing a beam that revolves around the optical axis at rotation frequencies of up to 19.4 Hz, corresponding to 7000 °/s. The setup was originally designed and assembled at Empa Thun, and two functionally identical replicas were subsequently built at Bern University of Applied Sciences (BFH) for process development and at the European Synchrotron Radiation Facility (ESRF, Grenoble) for operando high-speed X-ray imaging experiments. The optical path length, focusing conditions, and beam parameters were matched across the three installations and verified by beam profiling at each site, ensuring comparability between laboratory processing trials and synchrotron observations. Beam-profile measurements yielded a $LG_0^0$ spot diameter of approximately 66 µm at $1/e^2$, while the beam $HG_1^0$ exhibited an asymmetrical elliptical spot size of approximately 55 × 96 µm at $1/e^2$ for short and long axis measured with a (Basler acA1920-150µm) camera within the Rayleigh range.

## 2.2 In-situ synchrotron imaging during keyhole mode

In-situ synchrotron high speed X-ray imaging is used to directly observe keyhole dynamics and porosity formation movement in real time, providing mechanistic insight into the process transitions occurring across the parameter space. The goal of this study is also to gain insight into how SM can be used to affect keyhole and meltpool dynamics and the whole materials response across a whole beam spin with recording with 20kHz fps. All samples were placed on a custom holder attached to a 3-axis motorized stage and all laser scans were performed with 50mm/s velocity. The sample was irradiated in normal incidence under a flowing high-purity argon gas with flow rate of ∼2 L/min. All in situ synchrotron X-ray imaging experiments were carried out at the ESRF – The European Synchrotron (Grenoble, France) imaging beamline ID19. A polychromatic hard X-ray beam with a mean energy of ∼30 keV was used. Once the incident beam passes through the sample, the attenuated X-ray is converted by a LuAG: Ce scintillator and emits visible light. The visible light image is magnified by a 5 × objective and then captured by a Photron FASTCAM SA-Z 2100 K camera (Photron Ltd, Japan) with 20,000 images/s framerate. The field of view (FoV) was 2.25 mm × 2.25 mm with a pixel size of 2.2 µm. Each scan track was 3mm and was centered within the FoV of the X-ray image. The process dynamics during the onset, steady state, and end of laser scanning within the sample surface and subsurface were captured.

## 2.3 Image processing

All radiographs were processed in Fiji. Robust flat and dark references were obtained by taking the pixel-wise median of the flat- and dark-field stacks. Flat-field correction was then applied to each raw image I0 as flat field correction = (I0 − Dark)/(Flat − Dark), removing the fixed additive (dark current) and multiplicative (gain, illumination non-uniformity, scintillator/optic blemishes) components and leaving the sample transmission. Flagged pixels were replaced by the median of their valid 8-connected neighbors to remove residual stuck-pixel speckle, and corrected frames were stored as 32-bit floating-point images. The corrected images were denoised with VBM3D[22] and a custom background subtraction to suppress non-moving objects, after which the molten pool and droplet spatter were segmented by Otsu's method [23] and their geometries (length, width, area) quantified using built-in MATLAB functions.

The synchrotron videos have been segmented with the help of the model presented by Li et al.[24], which was finetuned with images coming from our dataset in order to improve the accuracy of the model. The segmentation divides each image in 3: the keyhole, the porosities and the background. The porosity area was measured in the upper and lower part of the meltpool separately (marked by the green and yellow rectangles in Supplementary Material Fig.S3. The keyhole depth was measured in the red rectangle (Supplementary Material Fig.S3) to prevent errors coming from wrongly segmented pixels at the edges of the full image.

## 2.4 Materials preparation and microstructural characterization

Three metallic materials were used in this study: 316L stainless steel, Ti6Al4V, and NdFeB. For 316L stainless steel, both 100 µm-thick foil and 1.5 mm-thick substrates were used, depending on the drilling and keyhole-melting experiments. Ti6Al4V coupons with dimensions of 40 × 10 × 2 $mm^3$ were used for machining/drilling trials. NdFeB samples with dimensions of 10 × 8 × 1 $mm^3$ were supplied by Webcraft AG, Switzerland, and were used for keyhole-melting and pore-dynamics investigations.

All samples were sectioned and prepared according to standard metallographic procedures. The sectioned specimens were cold mounted in a conductive resin using a Buehler Simplimet 4000 press. Surface preparation involved sequential grinding with silicon carbide (SiC) abrasive papers up to 2500 grit, followed by final mirror polishing using an oxide polishing suspension (OPS).

Microstructural characterization was performed using a TESCAN Mira scanning electron microscope (SEM) equipped with Electron Backscatter Diffraction (EBSD). EBSD measurements were conducted with a step size of 1 µm to obtain detailed crystallographic information. In addition, a Phenom series SEM (Thermo Fisher Scientific) was employed for imaging laser-drilled features and machined metallic parts.

A finite-volume model was developed in OpenFOAM v2506 to resolve the thermal and melt-flow behaviour under the two beam configurations. The model treated 316L stainless steel and air as an incompressible, laminar two-phase system using the volume-of-fluid method for interface capturing. Melting and solidification were described using an enthalpy-porosity formulation, while surface tension, thermocapillary flow, recoil pressure, radiative loss, and evaporative heat loss were included at the metal-air interface. The absorbed laser energy was imposed as a volumetric source, using a Gaussian distribution for the $LG_0^0$ beam and a rotating $HG_{10}$ distribution for the spinning-mode beam. The complete governing equations and model formulation are provided in Appendix A of the Supplementary Information.

# 3. Results

## 3.1 Rotational vectorial polarization filtering (RVPF)

RVPF was achieved using a free-space optical setup that converts the initial $LG_0^0$ laser beam into an $LG_1^0$ mode with either radial or azimuthal polarization; the polarization is then rotated by a λ/2 waveplate (spinning mode) or a λ/4 waveplate (profile-shifting mode) and finally filtered by a polarizing beam splitter or a Glan–Taylor polarizer. Fig. 1A shows the experimental configuration for the generation of the 2 drill-bit inspired beam modes. A linearly polarized fundamental ($LG_0^0$) beam propagates through (red beam path) motorized rotating half-wave plate (λ/2) (spinning mode) or by a rotating quarter-wave plate (λ/4) (profile shifting mode). The beam with rotating either linear or elliptical polarization then passes through an S-waveplate, which converts it into a vector mode with an annular ($LG_1^0$)profile, and a linear polarizer projects this vector field onto a two-lobe Hermite–Gaussian ($HG_1^0$)profile whose orientation is set by the local polarization. A plano-convex lens of f=150mm focuses the resulting field onto the substrate shown as the grey slab at the bottom.

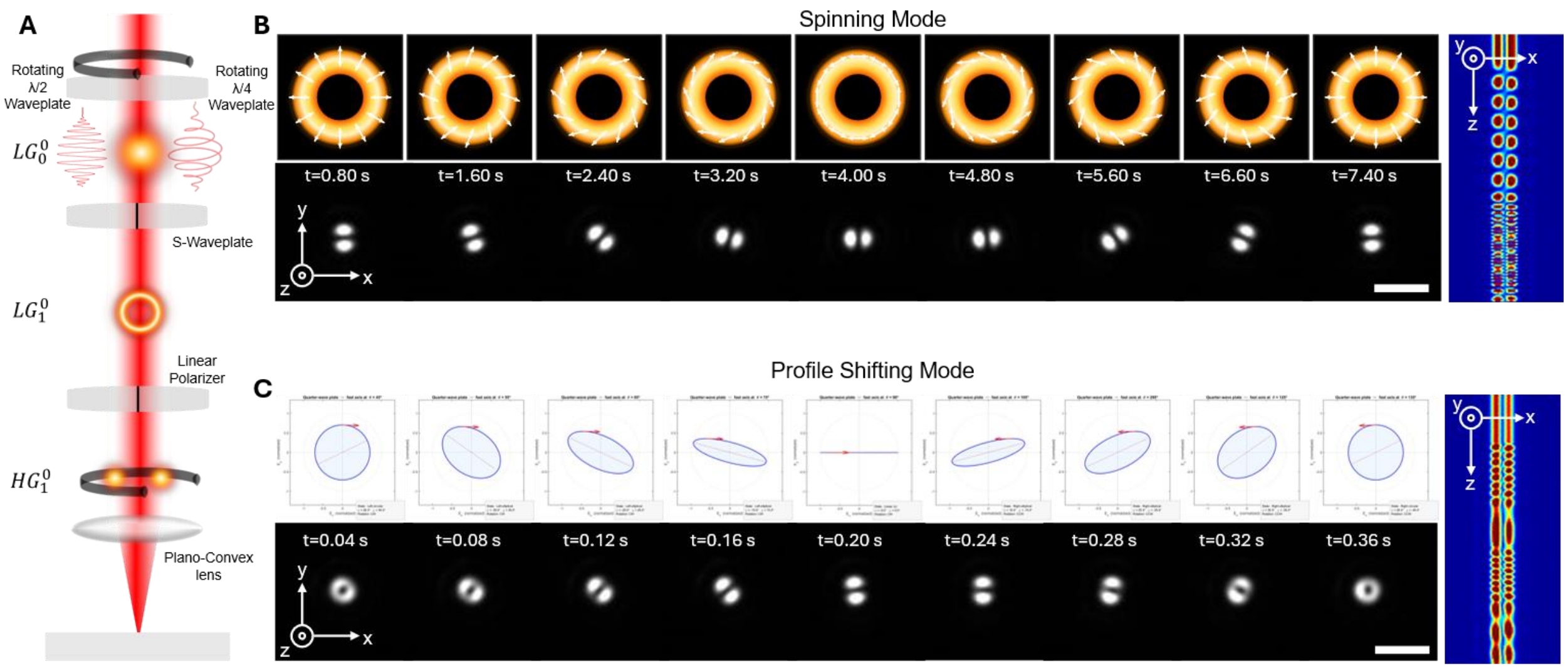

*Figure 1: Drill-bit beams generation and modes differentiation. Fundamental $LG_0^0$ beam conversion into an annular vector mode by a rotating half-/quarter-wave-plate pair, an S-waveplate, and a linear polarizer (A), then used in two modes:*

*continuous rotation of the focal profile by sweeping the polarization orientation "Spinning Mode," (B) and reshaping of the profile by sweeping the polarization ellipticity "Profile Shifting Mode,"(C) scale bar is 100μm.*

Fig. 1B, Spinning Mode (SM) shows the response when the polarization is rotated continuously by the rotating half-wave plate. The upper row shows the simulated vectorial polarization rotation caused by the half-waveplate spin for each time-frame while the bottom row is the transverse (x–y) laser spot images acquired over t = 0.80–7.40 s, acquiring a full $HG_1^0$ spin. The grayscale laser-spot images beneath, recorded simultaneously, start with the polarization parallel to the s-waveplate axis; each successive frame corresponds to a 23° rotation of the polarization and shows the $HG_1^0$ two-lobe profile spinning in lock-step with it, demonstrating continuous, deterministic rotation of the two lobes without instabilities (corresponding video in Supplementary Material, S1). The coordinate axes (lower left) define x and y in the transverse plane with z out of the page. The false-color image at the far right is the corresponding axial (x–z) projection, viewed with y into the page and z downward, whose periodic modulation along z reflects the focused-field structure. In this works laser processing results will focus only on SM configuration.

Figure. 1C shows the system response when the polarization ellipticity is swept by rotating the quarter-wave plate the beam profile shifting mode (PSM). The upper row presents the normalized polarization ellipse in the $(E_x, E_y)$plane at each step over $t = 0.04–0.36$ s, with the instantaneous field vector indicated by a red arrow. The polarization state evolves from circular to elliptical, then to linear, and back to circular with opposite handedness, spanning the full trajectory between the poles of the Poincaré sphere.

This behavior is analogous to an optical classifier that converts a two-lobe beam into an ideal doughnut-shaped profile, as observed at $t = 0.04$ s and $t = 0.36$ s, when the input polarization is perfectly circular. At intermediate times ($t = 0.08$ s and $t = 0.32$ s), a partially distorted state appears, where the ellipse approaches circularity but remains slightly imperfect. The grayscale images in the lower row are the laser spot at the focal point recorded and show the focal intensity profile evolving from an annular (doughnut) distribution under circular polarization, through a two-lobe structure under linear polarization, and back to an annular profile with opposite handedness. This confirms the spin-dependent shift of the intensity distribution. The false-color image on the far right shows the corresponding axial ($x–z$) projection, analogous to Fig. 1B. Notably, no discontinuity between spin states is observed, consistent with a continuous energy distribution. The beam retains a doughnut-like structure only under perfectly circular polarization passes the linear polarizer. A corresponding video of the PSM is provided in the Supporting Material (S2).

## 3.2 Enabling machining and drilling with CW radiation

The effect of the SM on machining and drilling is presented in Figure 2, where two material cases were investigated: drilling of thin 316L stainless-steel sheets and machining with consistent hole formation in 2mm-thick Ti6Al4V. All investigations were performed with the 1070 nm CW laser at a constant 10 W, measured for all beam types immediately before the focusing lens. The central comparison is laid out in Fig. 2A, which tracks drilling diameter as the focus is shifted relative to the surface of a 100-μm-thick 316L sheet for the right-handed SM beam (black, rotated at 5000 °/s) and a conventional $LG_0^0$ beam (red), each hole formed with a 2 s exposure. Additional top-view SEM images acquired at the same magnification are provided in the Supplementary Material, (Figs. S1 and S2), for the CW drilling evaluation of thin stainless steel foils.

Throughout the entire focus range, from −350 to +350 μm, the spinning beam carves holes of roughly 70–108 μm, while the $\mathrm{LG}_0^0$ beam manages only 9–21 μm just as importantly, the spinning-beam diameter barely falls away as the focus moves off the surface, revealing a far greater tolerance to focus position. The red curves trace the focusing caustic, with the focal spot sketched at its center, and each point is a mean with error bars representing the standard deviation of three independent measurements of the hole diameter for that condition (n = 3) and all irradiations were performed in ambient environment.

This robustness extends across beam types, as summarized in Fig. 2B, where the average drill diameter (Average D) measured on the sample with a 2 s exposure is plotted for a fundamental $LG_0^0$, a first-order Laguerre–Gaussian ($LG_1^0$), and a SM Hermite–Gaussian beam ($HG_1^0$) driven at 2000, 3000, 4000, and 5000 °/s. The two non-spinning modes drill small holes with a wide scatter ($LG_0^0$ ≈ 26 μm and $LG_1^0$ ≈ 38 μm, both carrying large error bars), whereas every spinning $HG_1^0$ condition lands consistently between roughly 90 and 100 μm with a much tighter spread, an outcome that proves largely insensitive to rotation speed over the range tested. As in Fig. 2A, points are means and error bars are the standard deviation of 3 independent diameter measurements per condition (n = 3).

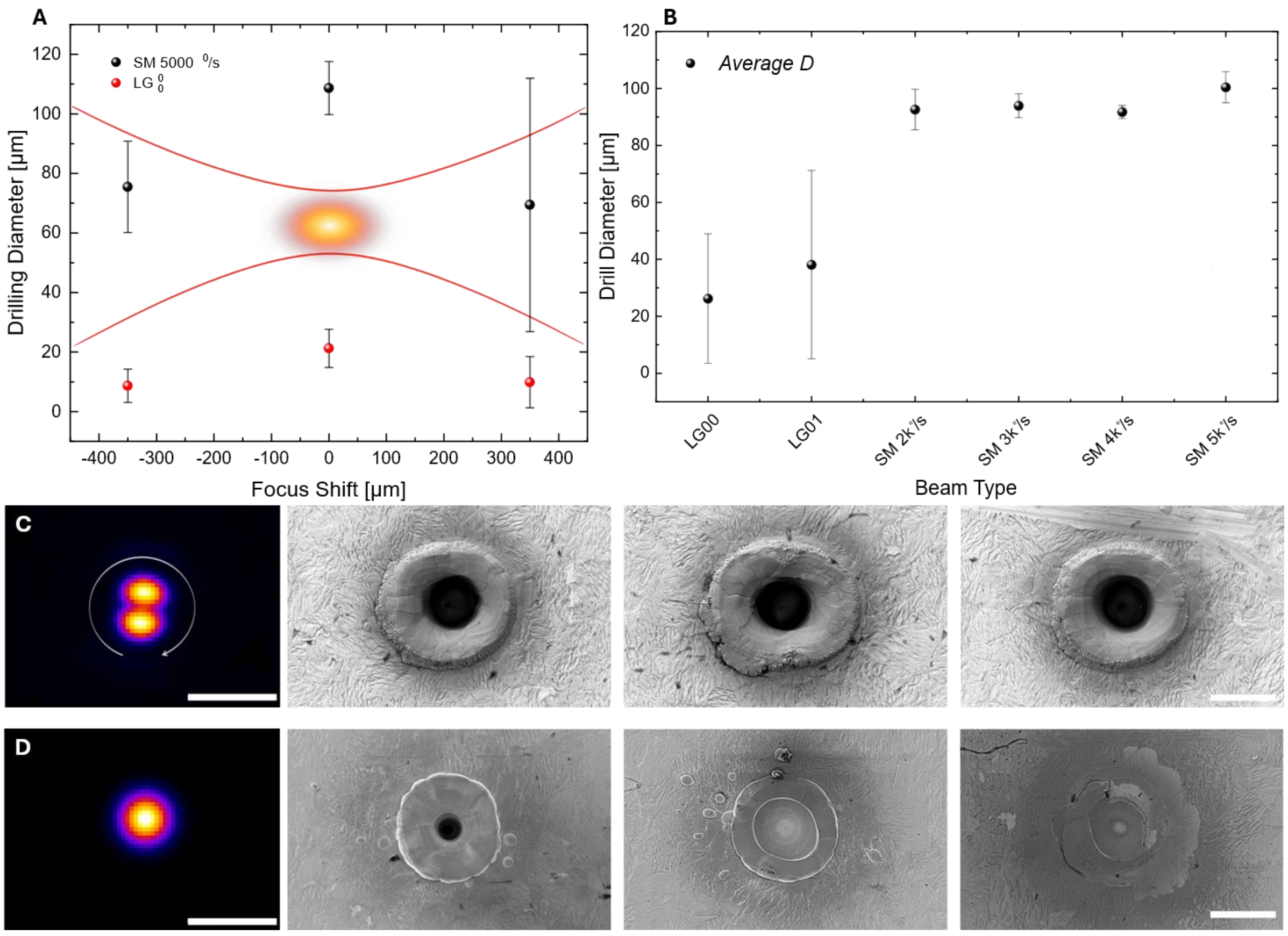

*Figure 2: A SM beam drills wider, more focus-tolerant holes than $LG_0^0$ or non-spinning modes. Drilling diameter versus focus shift for the spinning beam and a $LG_0^0$ beam (A) average drill diameter across beam types and rotation speeds (B) and focal intensity profiles with representative SEM images of representative holes for the spinning beam (C) and the $LG_0^0$ beam (D). Scale bars, 100 µm.*

The origin of this difference is made visible in Fig. 2C and Fig. 2D, which show representative holes machined repetitions through 2-mm-thick Ti6Al4V at a 2s exposure. In Fig. 2C, the false-color image at the left shows the measured focal intensity of the rotating two-lobe $HG_1^0$ mode, the white circular arrow marking its direction of spin, and the three SEM images alongside it present the resulting holes broad, consistently machined craters bounded by rough, raised rims. Fig. 2D provides the $\mathrm{LG}_0^0$ counterpart the single-lobe focal spot at the left and, beside it, SEM images of holes in the sample that are smaller and more sharply defined, with cleaner circular geometry. Together the two rows connect the spun, ring-like energy deposition of the structured beam to the wider, rougher openings it produces. Overall, these results indicate that the advantage of the spinning two-lobe beam is hydrodynamic rather than purely energetic. Whereas the static Gaussian beam deposits its energy at a fixed point relying solely on vaporization and recoil pressure to expel material, and allowing melt to stagnate and re-solidify at the hole walls the rotating intensity distribution sweeps an annular heating zone whose moving thermal gradient continuously shears the melt azimuthally. This sustained stirring drives molten material outward toward the rim, keeps the melt layer thin and mobile, and suppresses recast plugging, so that material is removed consistently over a wide, well-defined aperture.

## 3.3 Enabling enhanced convective flow and pore healing

Figures 3A and 3B show stitched image sequences extracted from operando X-ray recordings of keyhole-mode laser melting in 316L stainless steel. The recordings were acquired at a frame rate of 20 kHz, with the beam scanned at 50 mm $s^{-1}$, an average laser power of 170 W, a total acquisition time of 49.6 ms, and an approximate scan length of 3 mm. Figure 3A corresponds to the SM beam, whereas Fig. 3B corresponds to the $LG_0^0$ beam. The corresponding videos are provided in the Supplementary Material as Videos S3 and S4.

For the spinning beam, the keyhole depth exhibits a periodic variation associated with the rotation of the beam. This modulation is reflected in the porosity behavior, pore formation near the bottom of the melt pool increases when the keyhole reaches its maximum depth and decreases when the keyhole becomes shallower. In contrast, for the $LG_0^0$ beam, the keyhole depth remains comparatively constant over the analyzed sequence. However, porosity is generated over a broader range of melt-pool heights, and the amount of porosity in the upper part of the melt pool fluctuates significantly.

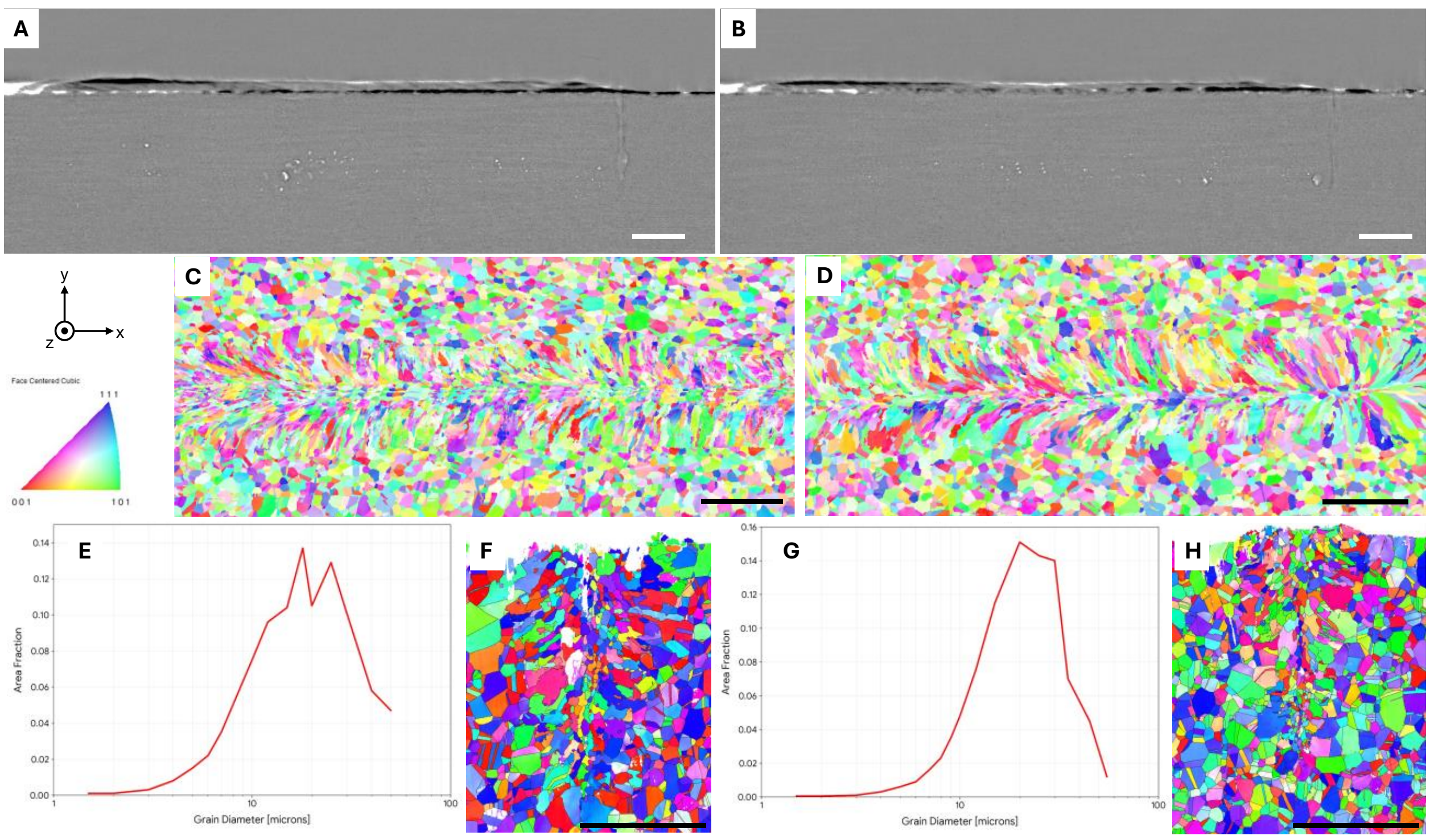

*Figure 3: Stitched images from operando X-ray imaging are shown for the (A) SM and (B) $LG_0^0$ beam. EBSD IPF maps are presented for (C) SM and (D) $LG_0^0$ beam conditions (IPF in the z direction (out of page)). Grain-diameter distributions are shown for (E) SM and (G) $LG_0^0$ beams. EBSDIPF [z] maps of the keyhole cross-sections are shown for (F) SM and (H) $LG_0^0$ beams. The scale bar is 200 μm.*

Fig.3C and Fig. 3D show crystallographic orientation maps obtained by electron backscatter diffraction (EBSD) for the solidified FCC 316L stainless steel processed with the spinning beam (Fig. 3C) and the LG00 beam (Fig. 3D), respectively. Grain orientations are represented by inverse pole figure (IPF) maps, in the z direction. The two maps are essentially indistinguishable as they both show the same fine, near-equiaxed grain matrix surrounding a central band of columnar, feather-like grains. This grain structure arises from epitaxial growth extending outward from the solidification line and exhibits grain size, texture, fusion-zone width, and grain-size distributions comparable to those observed in Fig. 3E and Fig. 3G, corresponding to the SM and $\mathrm{LG}_0^0$ beams respectively. This similarity is itself informative, as it confirms that the two beams melt equivalent volumes and therefore have closely matched spot sizes, both near the diffraction limit. Same similarity is also on the EBSD results on the meltpool cross-sections along z-axis presented at Fig.3F and Fig.3H for SM and $\mathrm{LG}_0^0$ beams respectively, the differences in keyhole stability and porosity in Fig. 3A,B therefore arise from the spinning dynamics of the beam rather than from any difference in spot size or deposited energy density.

With the spinning-mode beam, the keyhole depth was observed to vary periodically with beam rotation. Porosity formation at the keyhole bottom increased when the keyhole was deepest and decreased when it became shallower, as also observed (in the supplementary videos for steel S3 and S4). In contrast, porosity in the upper part of the melt pool remained very low and was mostly reabsorbed by the keyhole, indicating strong and stable convective flow induced by the shape and motion of the beam, even during the short irradiation time. This behavior is now highlighted more clearly in Fig. 4 for NdFeB case.

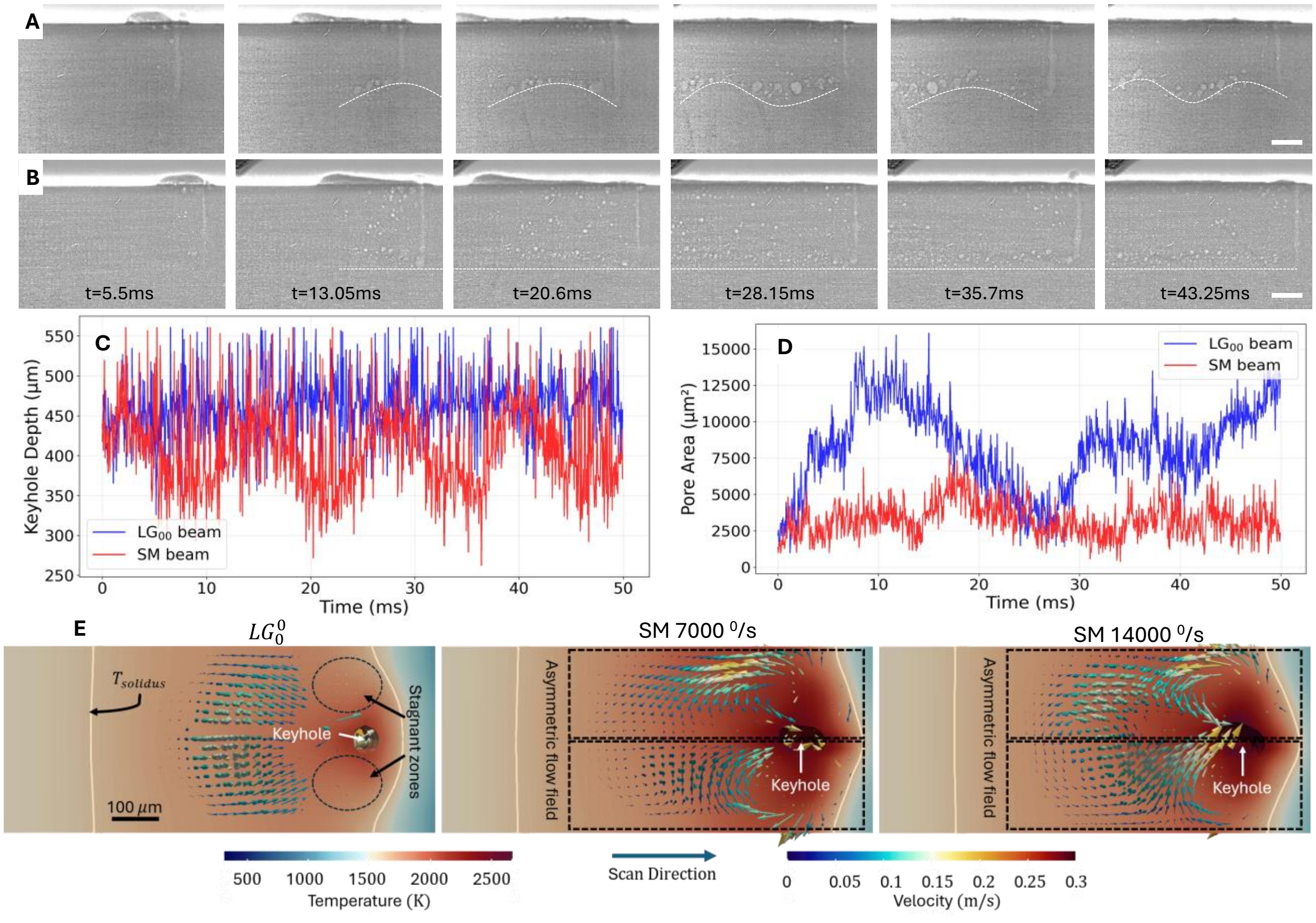

*Figure 4: Stitched images from operando X-ray imaging are shown for the (A) SM and (B) $LG_0^0$ beam. Keyhole depth as a function of time (C) and porosity area in the top half of the melt pool as a function of irradiation time (D), for the SM and $LG_0^0$ beam conditions. (E) Simulated velocity and temperature fields in the lateral cross-section at the mid-plane of the melt pool for the Gaussian beam and the $HG_{01}$ beam rotating at 7000 °/s and 14000 °/s. Scale bar, 200 µm.*

Figure 4 A,B presents operando X-ray recordings of laser keyhole regime in NdFeB and exactly the same irradiation parameters as discussed at Fig.3. The displayed at the matched times t = 5.5, 13.05, 20.6, 28.15, 35.7, and 43.25 ms beneath the columns Fig. 4A is the SM and Fig. 4B is the $\mathrm{LG}_0^0$ beam corresponding videos can be found in supporting material section (S5, S6). For the SM the delivered power was actively adjusted during rotation so that its time-averaged value matched that of the $\mathrm{LG}_0^0$ beam, ensuring that the two conditions are compared at equal average power. In each row the white overlay marks the trajectory of the keyhole tip: for the spinning beam it is essentially straight and level (Fig. 4A), indicating a stable keyhole, whereas for the $\mathrm{LG}_0^0$ beam it is strongly undulating (Fig. 4B), indicating an oscillating, unstable keyhole. This difference is mirrored in the porosity: in Fig. 4A the bright spherical features are gas pores that nucleate within the melt and then progressively shrink and disappear along the track, so that porosity is continuously healed as processing proceeds the one exception being near the tip, where the pores remain noticeably larger in Fig. 4B the pores instead persist throughout the melt and are distributed along the entire track rather than resorbed. A corresponding Supplementary Video (S7 and S8) shows that the power drops with $LG_0^0$ beam also suffers from pronounced porosity located all over the meltpool and well above the tip of the keyhole.

The contrast indicates that the SM supplies an additional driving force for melt transport: by continuously sweeping the energy around the pool, it both steadies the keyhole tip and sets up stronger, more symmetric Marangoni (thermocapillary) and convective flows that recirculate the melt and carry entrained bubbles back to the free surface, where they escape, healing the porosity. Because this is achieved at the same average power as the $LG_0^0$ beam, the improvement reflects how the energy is delivered rather than how much. The larger residual pores at the tip are attributed to a small instantaneous modulation of the power over each rotation cycle that persists despite matching of the average: the polarization is rotated through shaping optics whose transmission is not perfectly polarization-independent, so the instantaneous power ripples slightly as the polarization completes one full twist, and at the phase where it dips the convective stirring momentarily weakens, leaving bubbles insufficiently driven out and producing the larger pores at the tip. This residual ripple reflects imperfections in the current optics rather than an intrinsic limit of the method.

For the porosity analysis in (Fig.4C,D), the reported values correspond to the projected pore area within the selected region of interest, indicated by the green or yellow rectangle (supplementary Fig. S3 At the keyhole tip, the SM beam creates larger porosity than the $LG_0^0$ beam. However, the measured projected areas do not differ substantially, for two combined reasons. First, segmentation imperfections increase the apparent pore area in both datasets, reducing the contrast between the conditions. Second, the average pore size is larger for the SM beam near the tip, so a similar projected area can correspond to a larger pore volume when fewer but larger pores are present.

NdFeB has very low viscosity and surface tension in comparison with 316L steel[25], contains volatile constituents Nd has a high vapor pressure relative to Fe, and Nd loss by evaporation has been documented during reprocessing of sintered magnets[26,27]plus hydrogen/oxygen retained from the hydrogen-decrepitation and sintering route, since hydrogen desorption from HD-processed NdFeB occurs in several stages and is only complete at ~400 °C [28,29]. Low thermal conductivity (8–9 $W \cdot m^{-1} \cdot K^{-1}$ for sintered NdFeB [30] vs ~15–20 $W \cdot m^{-1} \cdot K^{-1}$ for 316L[31]) concentrates the absorbed energy, sustaining strong boiling and an unstable, violently fluctuating keyhole[32,33]. Keyhole-mode melting in LPBF NdFeB is known to produce extensive porosity and extremely brittle samples above a critical energy density[34,35] so gas/vapor bubbles are generated and shed throughout the melt pool, both above and below the keyhole [33,36], and rapid solidification traps them before they can escape [36,37]. Basically, it is an ideal material for observing melt fluid motion, since the pervasive pore population across the melt pool can act as tracer/indicator particles bubbles are routinely used in synchrotron X-ray imaging to track Marangoni-driven recirculation in melt pools [36,38]. In 316L, by contrast, the keyhole is comparatively stable at moderate normalized enthalpy, the melt is largely free of dissolved gases and volatile species, and higher surface tension and longer melt-pool lifetime allow most bubbles to coalesce and float out of the pool surface, so porosity is limited mainly to the classic keyhole-tip collapse mechanism, where vapor cavities pinched off at the keyhole tip are driven into the melt and trapped by the solidification front at the fusion-zone bottom.

Consequently, a finite-volume numerical model was also developed for 316L stainless steel to clarify the flow mechanisms under $LG_0^0$ and SM beam irradiation as presented at Fig.4E. The velocity fields extracted from the mid-plane cross-sections of the lower half of the melt pool reveal fundamentally different flow topologies. For the $LG_0^0$ beam, flow near the top surface is mainly confined to the immediate keyhole region, where surface-tension gradients drive radially outward flow from the hot keyhole rim toward the cooler melt-pool periphery. However, this active flow does not penetrate deeply into the melt-pool interior. Instead, the lower part of the pool remains largely stagnant because the radially symmetric, steady heat input organizes convection around the keyhole without generating strong circulation in the bulk melt. Pores trapped in this stagnant region experience little convective force and have no effective pathway back to the free surface or keyhole channel. As the beam translates, the active flow region moves forward, leaving pores behind in the stagnant wake, where they are eventually frozen in by the solidification front.

The SM beam produces a markedly different flow topology. Its non-axisymmetric double-lobe intensity profile continuously rotates, imposing time-dependent and spatially asymmetric thermal gradients on the melt-pool surface. This prevents the flow from becoming radially symmetric and steady, as in the $LG_0^0$ case. At any instant, the two lobes occupy non-equivalent positions relative to the scan direction, generating unequal Marangoni stresses on either side of the melt pool. Rather than cancelling, this broken symmetry reinforces a rotational circulation that draws fluid from the pool interior into the active flow field. Consequently, stagnant zones are reduced or eliminated.

Because of the beam rotation, the double-lobe intensity distribution repeatedly revisits fixed positions within the melt pool. A void-pore therefore experiences multiple interactions with high-velocity regions, rather than a single encounter as in the $LG_0^0$ case. Each interaction applies additional shear and displacement, increasing the probability that the pore is transported toward the keyhole or free surface and removed. The combined effects of reduced stagnant-zone volume and repeated pore re-engagement provide a physically consistent explanation for the experimental observation that pores remain persistent and immobile under $LG_0^0$ irradiation, but progressively move and disappear under spinning-beam irradiation.

## 3.4 Enabling surface texturing with extreme complexity

Figure 5 demonstrates the SM, implemented with RVPF, applied to ultrashort laser pulses. Fig. 5A shows linear scans of a conventional fundamental beam at scan speeds *u* increasing from right to left (0.05–0.15 m/s, in 0.02 m/s steps) these lines are indistinguishable from those of any static beam profile.

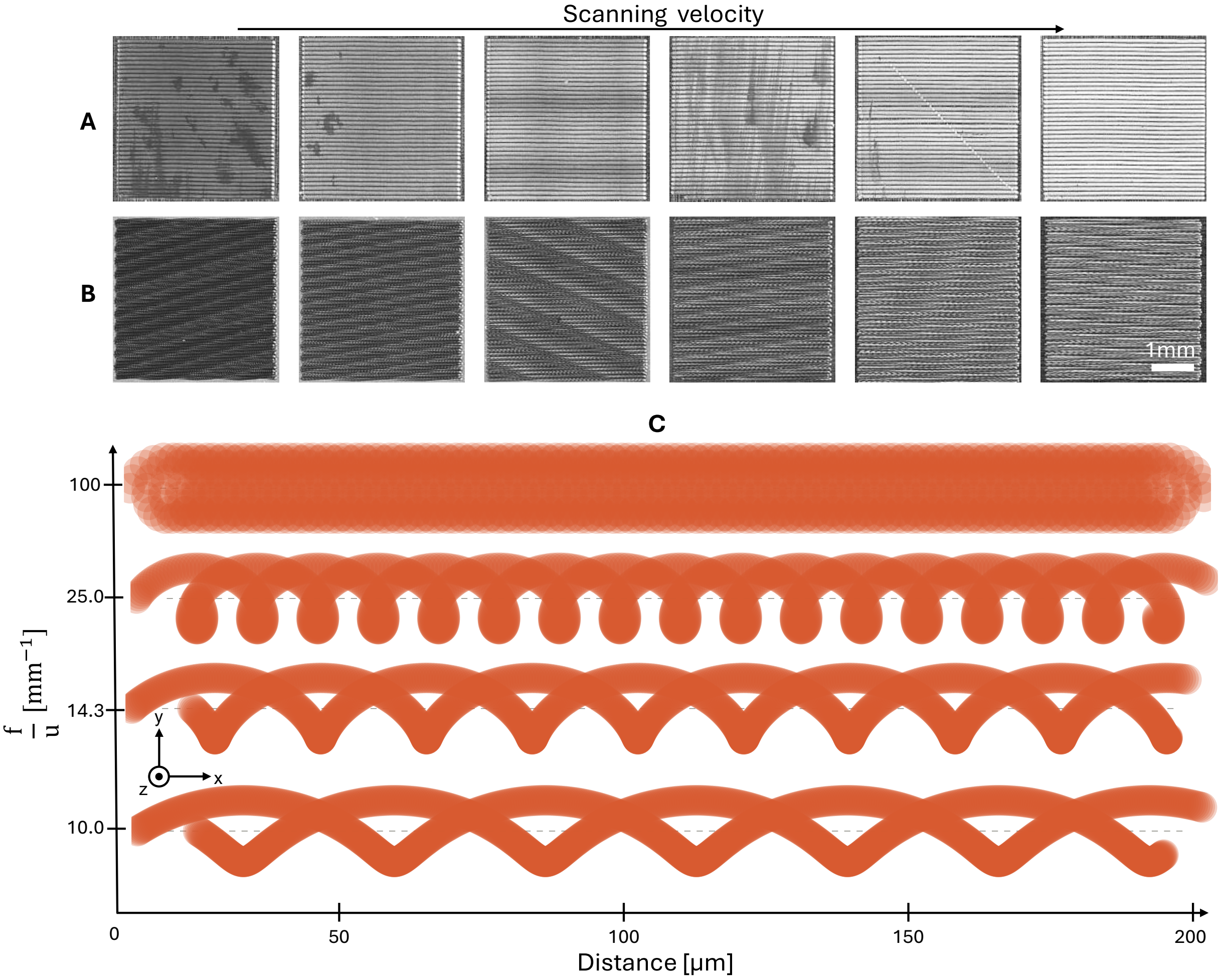

*Figure 5: Creating complex textures from simply linear line scans. (A) stitched images of a 4x4mm linear line scans with $LG_0^0$, (B) exact same scans but with right-handed SM beam 5000 °/s, (C) schematic illustration of $HG_1^0$ spinning at same frequency but with increasing speed every time. Line distance was 80um.*

Applying the same speeds and line spacings to the SM beam, spinning at 5000 °/s (≈ 13.9 Hz), instead yields a virtually unlimited family of woven, textile-like patterns, shown as stitched micrographs for the same speed series in Fig. 5B.

Such patterns can be designed deliberately and it is preferable to vary the SM frequency rather than *u* given that changing *u* alters the effective number of pulses per spot and hence the line thickness, complicating the design and simulation of the structures. The SM therefore provides an additional independent parameter for computationally designed surface textures and operated bidirectionally (meander) or quasi-statically under full automation can reach essentially any target texture with reduced processing time and added degrees of freedom, which is of clear value to the surface-texturing and marking industry. Moreover, representative photographs of the SM beam and $LG_0^0$ textured samples are provided in the Supplementary Material (Fig. S4), demonstrating the sensitivity of the resulting markings to key processing parameters, including the marking mode and other relevant marking conditions.

Fig. 5C plots the modelled track of the scanning, spinning beam as a function of the spin-to-scan speed ratio f/u ($mm^{-1}$) along the scan direction (distance 0–200 μm), for f/u = 100, 25.0, 14.3 and 10.0 $mm^{-1}$. Each trace is a single scan line rendered by the circle-stamping method, so the two intertwined strands are the paths of the two lobes and the stippled tone reproduces the accumulated-dose gradient, darkest where the strands cross and overlap. The morphology is governed by the along-scan pitch, $\Lambda = u/f = 1/(f/u)$: at f/u = 10 $mm^{-1}$ (≈ 100 μm pitch) the lobes form a wide, open braid with large lens-shaped gaps between crossings at 14.3 and 25.0 $mm^{-1}$ the period shortens and the loops tighten and at 100 $mm^{-1}$ successive passes overlap completely, merging into a continuous, uniformly filled band. Increasing f/u thus progressively converts the resolved two-lobe weave into a homogeneous track. Superimposed on this pitch dependence is a handedness-induced asymmetry: because the rotation direction combines with the scan vector, the net velocity of each lobe over the surface and hence its local dwell time and deposited dose differs on the two sides of the nominal scan line. For a right-handed spin the dose is concentrated toward the upper side (+y) of the line, whereas a left-handed spin shifts the heavier dose to the lower side the asymmetry therefore reverses with the spin handedness relative to the scan direction and vanishes only for a non-spinning (symmetric) beam.

## 4. Dose metrics for the spinning HG01 (SM) beam (CW & Pulsed)

To quantify how the deposited dose is distributed under the SM mode expressed as the effective number of pulses per spot for pulsed lasers, or as the energy/power dose per spot area for CW sources we first formulate the corresponding mathematical expressions. Unlike a $\mathrm{LG}_0^0$ beam, an $HG_{01}$ ($TEM_{01}$) mode has a two-lobe intensity profile separated by a dark central node, so when the beam is simultaneously scanned at speed u and spun about its propagation axis at frequency f, the dose received at a given point is governed by the ratio of the spin frequency to the scan speed. The expressions below describe the fast-spin (high-f/u) limit, in which a surface point samples all azimuthal phases the resulting effective profile is symmetric in *Y* and therefore does not capture the handedness-induced dose asymmetry of 3.4, which is a finite-f/u effect.

In the following we derive the single-pulse (single-pass) distribution, the spin-averaged effective profile, and the resulting effective pulse number for pulsed operation and energy dose for CW operation.

For pulsed operation, the fluence accumulated on the surface along a single line scan is independent of the along-scan coordinate:

$$F_{acc}(Y) = \frac{f_{rep}\, E_p}{u\, w_0\, \sqrt{2\pi}} \left(1 + \frac{4\, Y^2}{{w_0}^2}\right) exp\left(\frac{-2\, Y^2}{{w_0}^2}\right) \tag{1}$$

peaking not on the axis but on two ridges at Y = $\pm w_0/2$ (Figure 6), where $F_{acc,max} = \sqrt{\frac{2}{\pi}}\, e^{-1/2}\, \frac{f_{rep}\, E_p}{u\, w_0} \approx 0.484\, \frac{f_{rep}\, E_p}{u\, w_0}$. Normalizing to the single-pulse lobe peak $F_0 = \frac{4\, E_p}{\pi\, e\, {w_0}^2}$ gives the effective number of overlapping pulses,

$$N_{eff}(0) = \frac{F_{acc}(0)}{F_0} = \left(\frac{e}{4}\right) \sqrt{\frac{\pi}{2}}\, \frac{f_{rep}\, w_0}{u} \approx 0.85\, \frac{f_{rep}\, w_0}{u} \tag{2}$$

For CW operation the same results hold under the substitution $f_{rep} E_p \rightarrow P$ (P the average power): the energy dose is F(Y) with peak $F_{max} \approx 0.484\, \frac{P}{u\, w_0}$, the energy deposited per spot is $E_{spot} = P\, t_{eff}$ via the effective dwell time $t_{eff} = \left(\frac{e}{4}\right) \sqrt{\frac{\pi}{2}}\, \frac{w_0}{u}$ (so that $N_{eff} = f_{rep}\, t_{eff}$), and the mean areal energy dose for a raster of hatch spacing h is

$$D_A = \frac{P}{u\, h} \tag{3}$$

Where $F_{acc}$, accumulated fluence ($J \cdot cm^{-2}$) Y, transverse distance from the scan line $f_{rep}$, pulse repetition rate $E_p$, pulse energy P, average power u, scan speed $w_0$, $LG_0^0$ waist ($1/e^2$ radius) f, spin frequency $F_0$, single-pulse lobe-peak fluence $t_{eff}$, effective dwell time $E_{spot}$, energy per spot h, hatch spacing $D_A$, areal energy dose ($J \cdot cm^{-2}$) $N_{eff}$, effective pulse number.

Figure 6 provides a spatial complement to Eq. (1)–(3), making explicit how the dose of the spinning SM beam accumulates along a single line scan.

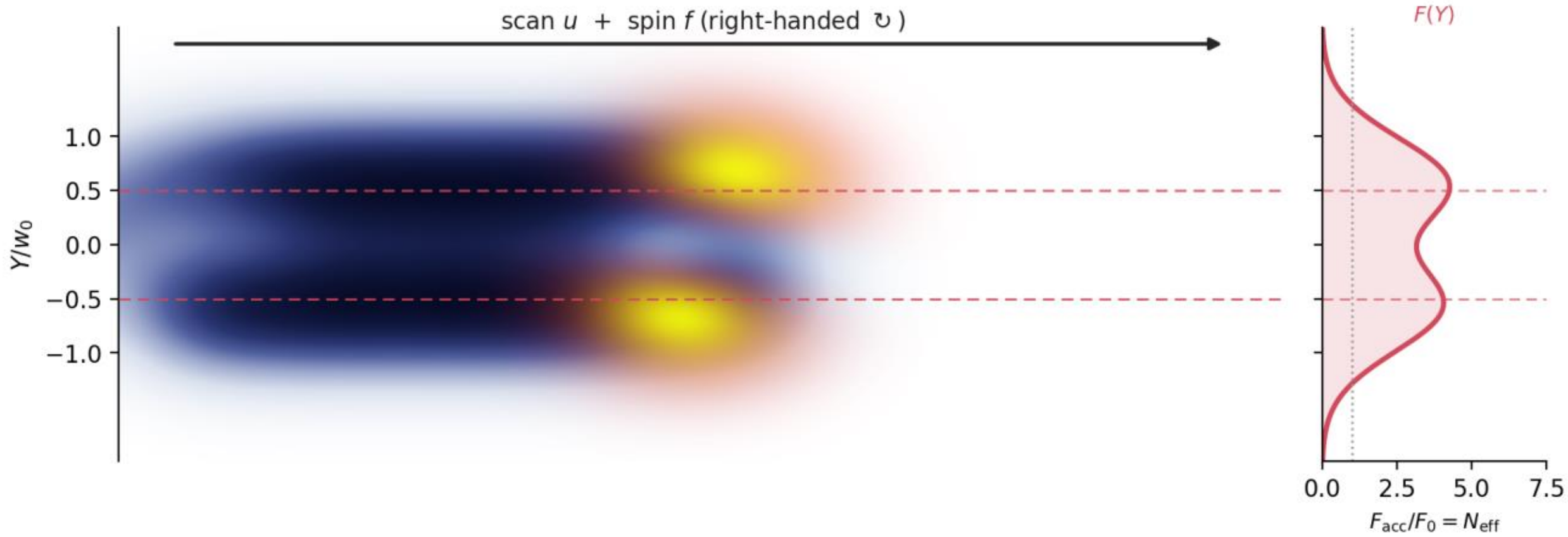

*Figure 6: Accumulated dose of a right-handed SM beam. The dark-blue gradient gives Neff = Facc/$F_0$, converging to twin ridges at Y = ±$w_0$/2 with a shallow central node.*

As the two-lobe footprint is simultaneously translated and spun successive imprints overlap and build up in the figure this accumulation is rendered as increasing darkness, which is directly proportional to the effective number of overlapping pulses, $N_{eff} = F_{acc}/F_0$. The converged transverse profile is not a single on-axis track but two symmetric ridges at $Y = \pm w_0/2$separated by a shallow central node, whose dose is $e^{1/2}/2 \approx 0.82$of the ridge value a partially filled minimum rather than a true null, because the scan line still integrates through both bright walls of the spin-averaged annulus. This near-plateau, twin-ridge distribution accounts for the broad, near-uniformly dosed tracks characteristic of the SM beam, whose width and fluence can be tuned independently through the spin-to-scan ratio Eq. (2) the additional processing degree of freedom exploited in 3.4. We emphasize that Figure 6 represents the fast-spin (high-$f/u$) limit, in which a surface point samples all azimuthal phases so that the effective profile is azimuthally averaged and therefore symmetric. It actually corresponds to the homogeneous-track limit toward which the resolved, handedness-asymmetric morphologies of Fig. 5C converge as $f/u$ increases.

## 5. Conclusions

In conclusion, we have presented a scalable, drill-bit-inspired dynamic beam-spinning approach to laser materials processing that opens a range of new possibilities across the broad field of laser materials processing and can serve as an additional, independent processing parameter capable of advancing the field. Specifically, we introduced rotational vectorial polarization filtering (RVPF) for realizing the SM and PSM configurations, and demonstrated that consistent machining is now achievable even with low-power CW sources with further technological development higher spin frequencies, or the implementation of purely optical, ultrafast spinning we anticipate that material ablation with such beams may also become feasible. Using synchrotron imaging, we further presented recording and images of the enhanced melt flow induced by the spinning beam, which could benefit pore healing or resorption and material mixing for additive manufacturing. Finally, we showed that the method enables the fabrication of complex, sophisticated, and richly tunable programmable surface textures from simple linear scans. We regard these results as only a starting point and expect that much more will emerge from this method across laser manufacturing, optical data storage, laser-based lithography, and microscopy.

## Acknowledgements

The authors thank Mr. Christoph Amsler and Mr. Tomas Rytz for their technical support throughout the project, with laser commissioning and the custom machining of parts. The authors also gratefully acknowledge the European Synchrotron Radiation Facility (ESRF) for the provision of beamtime, and the technical staff of beamline ID19 for their assistance during the in-situ X-ray imaging experiments.

## Conflict of interest

The authors declare no conflict of interest.

## Author contributions

E.S. conceived the project and invented the process and concept, designed and built the optical setup, designed and performed the experiments and the characterization, and wrote the manuscript. P.Y. contributed to the synchrotron X-ray experiments and performed the material preparation and characterization. J.H. performed the simulations. A.W. performed the data processing. X.M. performed the EBSD measurements. A.R. supported the synchrotron X-ray imaging experiments. R.K., B.N., P.H., E.I. and S.S. provided resources. P.Y., A.W., X.M. and S.S. reviewed and edited the manuscript. All authors discussed the results and commented on the manuscript.